\documentclass[10pt,paperpaper]{article}
\usepackage[space]{cite}
\usepackage{opex3}
\usepackage{bm}
\usepackage{mathrsfs}
\usepackage{amssymb,amsmath}

\begin{document}

\title{Observation of non-diffracting behavior at the single-photon level}

\author{H\'ector Cruz-Ram\'irez,$^{1}$  Roberto Ram\'irez-Alarc\'on,$^{1,2}$ Francisco J.  Morelos, $^{1}$ Pedro A. Quinto-Su,$^{1}$ Julio C. Guti\'{e}rrez-Vega,$^{3}$ and Alfred B. U'Ren$^{1*}$}

\address{$^1$Instituto de Ciencias Nucleares, Universidad Nacional
Aut\'onoma de M\'exico, apdo. postal 70-543, 04510 D.F.,  M\'exico\\
$^2$Divisi\'on de Ciencias e Ingenier\'ia, Universidad de
Guanajuato, Loma del Bosque No. 103 Col. Lomas del Campestre C.P
37150 A.Postal E-143 Le\'on, Guanajuato, M\'exico\\
$^3$Photonics and Mathematical Optics Group, Tecnol\'ogico de Monterrey, Monterrey 64849, M\'exico }

\email{$^*$alfred.uren@nucleares.unam.mx} 



\begin{abstract}
We demonstrate the generation of non-diffracting heralded single photons,   i.e. which are characterized by a single-photon transverse intensity distribution which
remains essentially unchanged over a significant propagation distance.   For this purpose we have relied on the process of spontaneous parametric downconversion (SPDC) for the generation of signal and idler photon pairs, where our SPDC crystal is pumped
by a Bessel-Gauss (BG) beam.    Our experiment shows that the well-understood non-diffracting behavior of a BG beam may be directly mapped to the signal-mode,  single photons heralded by the detection of a single idler photon.  In our experiment, the heralded single photon is thus arranged to be non-diffracting without the need for projecting its single-photon transverse amplitude, post-generation, in any manner.
\end{abstract}



\section{Introduction}

Diffraction is ubiquitous in any branch of physics that involves waves, including  optics, acoustics, and quantum mechanics. For many applications, diffraction is a limitation and several techniques have been proposed for its control. Non-diffracting beam (NDB) is the common term used in classical optics to refer to a solution of the wave equation that propagates along a given axis preserving its transverse intensity pattern. The best known NDB is the Bessel beam introduced by Durnin et al. in 1987~\cite{durnin87}. Bessel beams were followed by the discovery of other nondiffracting solutions including Mathieu beams, parabolic beams, X-waves, and Airy beams~\cite{turunen09}.  Ideal NDBs have an infinite extent and energy, and thus they are not physically realizable. In view of this, high-quality apodized versions of ideal NDB's have been experimentally generated using annular masks, axicons, laser cavities, holograms and light modulators~\cite{gutierrez05}.

In the process of spontaneous parametric downconversion (SPDC)~\cite{burnham70}, individual photons from a pump beam which illuminates a second-order nonlinear crystal may  be annihilated with a small probability, each annihilation event leading to the emission of a signal and idler photon pair.    The detection of the idler photon of a given pair may ``herald'' the existence of a single photon in the signal mode; this is then known as a heralded single photon source.  SPDC is remarkable in that the photon pair properties can be widely engineered to fit particular needs~\cite{torres11}.  For a sufficiently short crystal, the two-photon state depends solely on the pump, with the resulting spectral (spatial) entanglement highly dependent on the pump spectral~\cite{uren03} (transverse $k$-vector~\cite{walborn10}) amplitude;  these degrees of freedom are in fact linked, e.g. the spatial pump properties may influence the photon-pair spectral properties~\cite{valencia08}.   Past works which exploit  SPDC together with transversely structured beams, of which NDB's are a sub-category, include the generation of entangled states in modes, e.g. Gauss-Laguerre, with orbital angular momentum~\cite{molinaterriza07}.   In many of these experiments, the signal and idler fields are projected onto specific transverse $k$-vector amplitudes using phase holograms or other optical elements.   In this paper we exploit the fact that for a sufficiently-short crystal, the spatial pump beam properties can be \emph{directly} mapped to the two-photon state~\cite{monken98, molina05}, and in particular to the heralded single-photon; this is accomplished without the need  for projecting, post-generation,  the heralded single photon into particular $k$-vector distributions~\cite{lekki04}.   Specifically, in our work we directly map the non-diffracting behavior of a Bessel-Gauss (BG) pump beam to the single-photon amplitude of the heralded signal photons.    Thus, we take the well understood non-diffracting properties of BG beams into the quantum optics realm.   
 
The ability to control the transverse spatial intensity pattern at the single photon level, while also achieving non-diffracting behavior, is notable.  Applications could be found in the area of free-space quantum communications, where the ability of an NDB to maintain its form despite the presence of obstacles~\cite{bouchal98} and/or despite propagation through a turbulent medium~\cite{gbur07, noriega07}. would be particularly useful.  Applications could likewise be found in  the controlled interaction of single photons with atoms and/or ions in the  linear arrangement of an ion trap or an optical lattice.  

\section{Theory}

Our main objective in this paper is the generation of non-diffracting single-photon wavepackets.  For this purpose, we take advantage of the well-developed field of classical NDB's.  

We rely on the process of SPDC to generate signal and idler photon pairs
in a bulk second-order crystal pumped by a BG beam of frequency $\omega_p$. 
We detect an idler photon on the Fourier plane defined by an $f$-$f$ optical system following the crystal, which heralds a
conjugate signal-photon wavepacket.  Under ideal conditions the quantum state of the signal photons, conditioned by the detection of an idler photon of frequency $\tilde{\omega}_i$ and transverse wavevector $\tilde{\textbf{k}}_i$, may be written as follows

\begin{equation}\label{E:state}
|\Psi\rangle_s=\int d \textbf{k}^\bot_s S_p(\textbf{k}^\bot_s+  \tilde{\textbf{k}}^\bot_i) | \omega_p-\tilde{\omega}_i,\textbf{k}^\bot_s \rangle,
\end{equation}

\noindent in terms of the angular amplitude $S_p(\textbf{k}^\bot)$ of the pump, and where $|\omega,\textbf{k}^\bot \rangle$ is
a single-photon Fock state of frequency $\omega$ and transverse wavevector $\textbf{k}^\bot$.

Here, ideal  conditions refers first to the use of a sufficiently short crystal and second to the use of a conditioning
idler-mode detector with sufficiently small transverse dimensions.   Indeed, in general the single-photon amplitude in Eq.~(\ref{E:state}) should be  multiplied
by a sinc function factor, determined by crystal properties such as crystal length and Poynting vector walk-off; for a sufficiently
short crystal, the width of this factor is much larger than that of the function $|S_p(\textbf{k}^\bot_{s}+\tilde{\textbf{k}}^\bot_{i})|$, so that the former may be disregarded~\cite{ramirez-alarcon12}.
If the idler detector has a non-zero transverse extent, the signal-mode single-photon state becomes a statistical mixture of states of the form in  Eq.~(\ref{E:state})
integrated  over all $\tilde{\textbf{k}}^\bot_{ i}$
values within the angular acceptance of the idler detector.  Equation~(\ref{E:state}) already contains the main physical mechanism which we exploit in this paper:  the signal-mode, single-photon amplitude corresponds to the
angular amplitude of the pump mode, except displaced in transverse momentum space.     Thus, our experiment  (to be described in detail below)  
has been designed 
so that, ideally, the detection of a single idler photon heralds a single signal-mode photon with a transverse amplitude which corresponds
to a direct mapping of the pump transverse amplitude.

Note that for a plane-wave pump, with $|S_p(\textbf{k}^\bot)|^2=\delta(\textbf{k}^\bot)$, the above mechanism implies that the signal-mode conditional angular spectrum
is likewise a delta function $\delta(\textbf{k}^\bot_s+  \tilde{\textbf{k}}^\bot_i)$.   In this paper our pump beam has a non-zero transverse-momentum width.    In particular, it has a BG transverse amplitude which is mapped to the conditionally-prepared single photons.   Because BG beams exhibit a non-diffracting character, this behavior will likewise be presented by the signal-mode single photons conditioned by the detection of an idler photon.

A BG beam is a conical 
coherent superposition of Gaussian beams, each with a radius at the beamwaist parameter $w_0$, and with a cone
opening half-angle  $\mbox{arcsin}(k_t/k_p)$, where $k_t$ is transverse wavenumber and $k_p$ is the pump wavenumber.  For a BG pump, the function
$S_p(\textbf{k}^\bot)$  may then be written as

\begin{equation}\label{E:BGampl-k}
S_p(\textbf{k}^\bot)= A \exp\left(-\frac{w_0^2}{4} |\textbf{k}_\bot|^2\right) I_0\left(\frac{k_t w_0^2 |\textbf{k}_\bot|}{2}\right),
\end{equation}

\noindent where $A$ is a normalization constant and $I_0(.)$ is a zeroth order
modified Bessel function of the first kind.   When viewed in the transverse position $\boldsymbol{\rho}$ domain, the transverse amplitude as a function of the propagation distance $z$ becomes  

\begin{equation}\label{E:BGampl-x}
A_p( \boldsymbol{\rho}) =A'\frac{1}{\mu} \exp \left\{-\frac{1}{\mu} \left( \frac{i k_t^2 z}{2 k_p }  +\frac{ |   \boldsymbol{\rho}   |^2}{ w_0^2} \right)  \right\} J_0\left(\frac{k_t |\boldsymbol{\rho}|}{\mu}\right),
\end{equation}

\noindent  in terms of $\mu=1+i z/z_r$ with $z_r=k_p w_0^2/2$ the Rayleigh range of the pump, the zeroth-order Bessel function  $J_0(.)$ and a normalization constant $A'$ .   A measure of the propagation distance over which the  single-photon transverse intensity pattern remains unchanged is given by  $z_{max}=w_0 k_s/k_t$,  where $k_s$ is the signal wavenumber.

\section{Experiment}

\begin{figure}[ht]
\centering
\includegraphics[width=8cm]{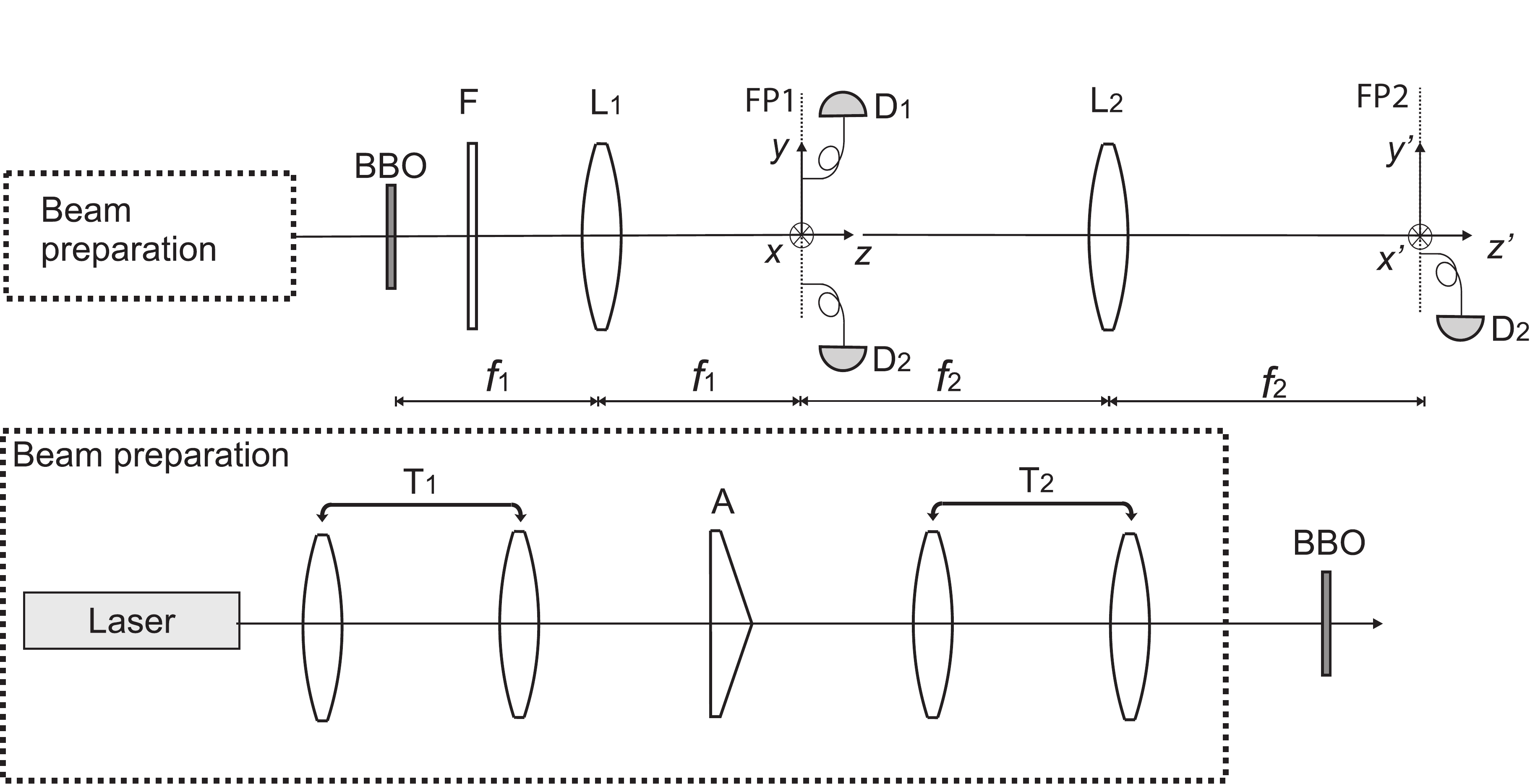}
\caption{ Experimental setup for the generation of non-diffracting single photons.}\label{Fig:setup}
\end{figure}

We have prepared a BG beam
to be used as pump in a second-order nonlinear crystal, which generates 
photon pairs through the SPDC process. Figure~\ref{Fig:setup} shows our experimental setup.  
A beam from a diode laser, centered at $406$nm with  
$72$mW power is sent through a telescope (T1) with magnification $\times 13.3$ to obtain
an approximately Gaussian beam with an approximate radius at the beamwaist of $7.5$mm.   This magnified beam is transmitted
through an axicon (A), i.e. a conical lens, with a $2^\circ$ apex angle, which maps the incoming beam into a high-quality BG beam.  As a final preparation step, this beam is magnified $\times5$ with a telescope (T2)
consisting of appropriately separated lenses with focal lengths $10$cm and $50$cm.

\begin{figure}[ht]
\centering
\includegraphics[width=8cm]{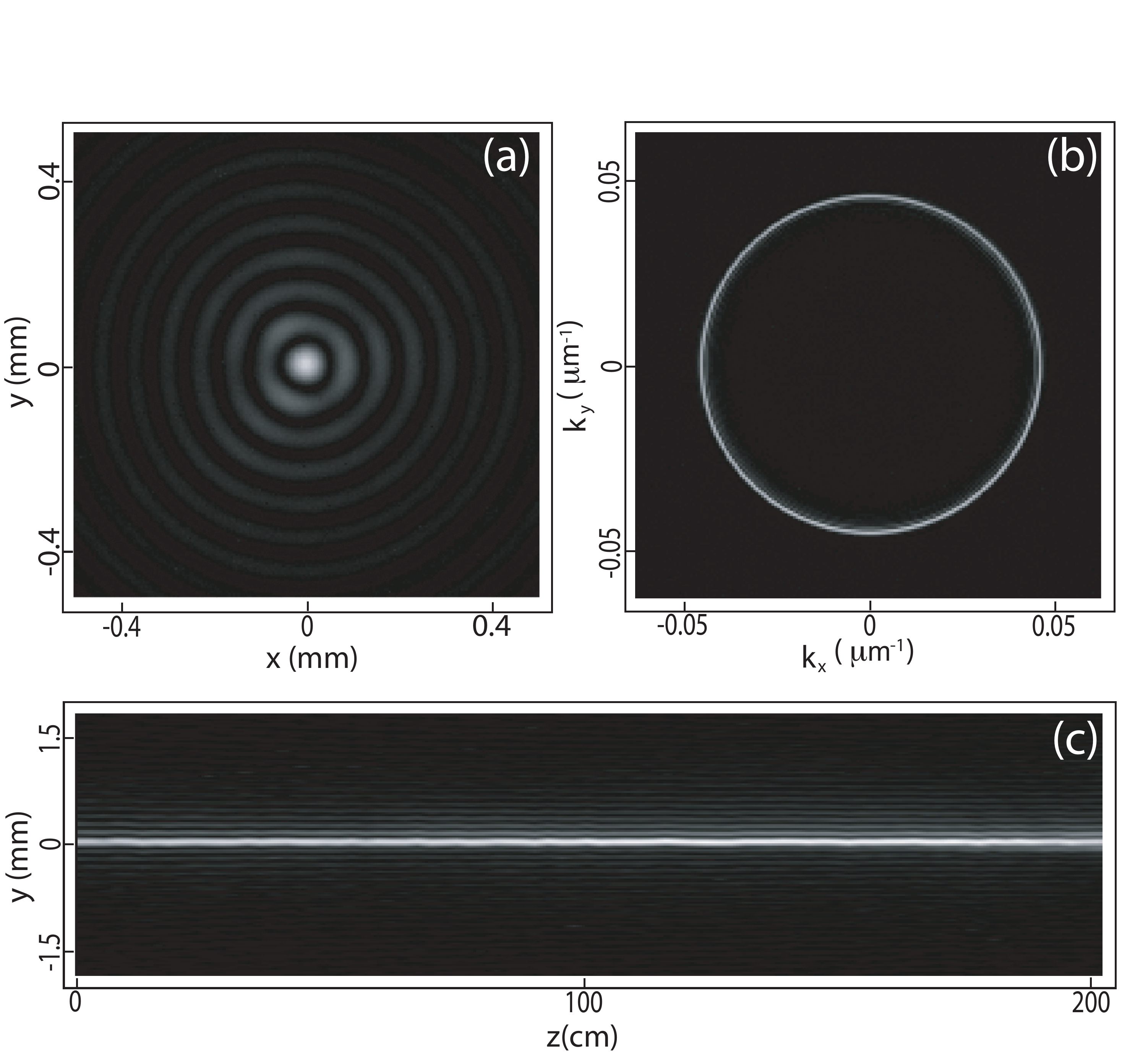}
\caption{BG pump properties: a) Measured transverse intensity, b) measured angular spectrum, and c)  measured transverse intensity vs coordinate $y$, measured under propagation along coordinate $z$.}\label{Fig:pump}
\end{figure}

Figure~\ref{Fig:pump}(a) shows a measurement of the transverse intensity distribution of the BG pump, obtained with a CCD camera, at a distance of $25$cm following the second 
lens of telescope T2 (see Fig.~\ref{Fig:setup}); as discussed below, this corresponds to the plane where the nonlinear crystal is placed.    This measurement
shows the distinct characteristics of a BG beam, i.e. a central spot surrounded by a set of concentric rings, as predicted by $|A_p(\boldsymbol{\rho})|^2$ in Eq.~\ref{E:BGampl-x}. The slight departure from azimuthal symmetry is probably
due to fabrication imperfections in the axicon.  Figure~\ref{Fig:pump}(b) shows
the corresponding transverse momentum distribution, obtained through an optical $f$-$f$ system prior to the CCD camera.    This figure shows the expected
annular shape of the BG pump angular spectrum as predicted by the function $|S_p(\textbf{k}^\bot)|^2$ in Eq.~\ref{E:BGampl-k}.  Note that the radius of this annulus,
in $\textbf{k}^\bot$ space directly yields the parameter $k_t$ of Eq.~(\ref{E:BGampl-k}), in our case with a value of  $k_t=0.046 \pm .001 \mu \mbox{m}^{-1}$.  Note also that the
$w_0$ parameter in Eq.~(\ref{E:BGampl-k}) may be obtained from the width of the annulus $\delta k$ according to the relationship $w_0=4/\delta k$; in our case we obtain a value of
$w_0=1.85 \pm 0.60$mm.

In order to visualize the propagation properties of the BG pump beam, Fig.~\ref{Fig:pump}(c) shows a plot of the measured transverse pump intensity distribution along the $y$ direction,
measured with a CCD camera, as a function of the propagation distance $z$.   It may be appreciated that this distribution remains essentially unchanged in a distance of over $200$cm, making it clear that our pump
beam has non-diffracting properties.   

\begin{figure}[ht]
\centering
\includegraphics[width=8cm]{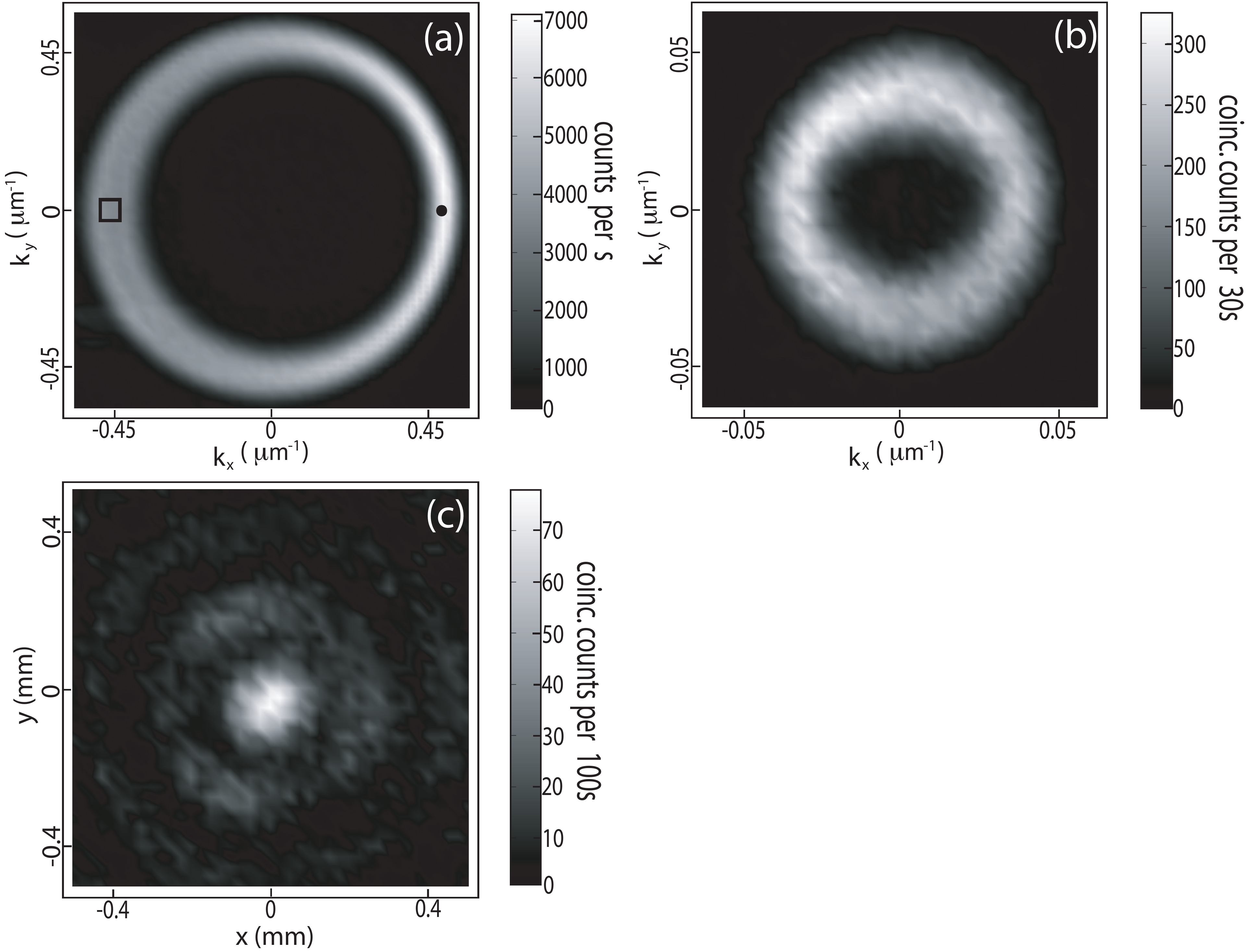}
\caption{  a) Measured angular spectrum of SPDC photon pairs.  b) Measured conditional angular spectrum of signal-mode heralded single photons. c) measured  transverse intensity of signal-mode heralded single photons. }\label{Fig:z0}
\end{figure}

The SPDC crystal, $\beta$-barium borate (BBO) with thickness  $1$mm, is placed a distance
of $25$cm from the second lens of telescope T2.  Pump photons are eliminated using appropriate filters (F).
An  $f$-$f$ optical system is used in order to yield a Fourier plane on which we can probe the signal and idler
transverse momentum distributions.   Specifically, a lens (focal length $f_1=10$cm and $1$-inch diameter; L1)  which is placed at a distance of $10$cm 
from the crystal, defines a Fourier plane (FP1) at a distance of $10$cm from the lens.     Spatially-resolved photon counting on this plane
yields the angular spectrum of the SPDC photon pairs.   For this purpose, we have used the fiber tip of a large-diameter fiber ($200\mu$m) which can be 
displaced laterally along the $x$ and $y$ directions with the help of a computer-controlled motor ($50$nm resolution and $1.5$cm travel range).  The fiber used leads
to a Si avalanche photodiode (APD), with its output connected  to standard pulse-counting equipment to obtain number of detection events per second data; Fig.~\ref{Fig:z0}(a)
shows the resulting measured SPDC angular spectrum.

Note that the annular SPDC angular spectrum is asymmetric in a manner similar to that
observed for a focused Gaussian pump beam (e.g. see Ref.~\cite{bennink06}).  The distinct properties resulting 
from a BG beam pump are, however, more evident when we analyze spatially-resolved coincidence counts.    For this purpose, we
use spatially-resolved coincidence photon counting implemented with \emph{two} independent fiber tips, each laterally displaceable
with a computer-controlled motor and each leading to an APD.

Figure~\ref{Fig:z0}(b) shows the measured conditional
angular spectrum of the signal photon, i.e.  conditioned to the detection of an idler photon with a fixed transverse momentum value.  Specifically, the idler fiber tip is placed at the disk indicated in Fig.~\ref{Fig:z0}(a), corresponding to $k^\bot_{iy}=0$ and with a $k^\bot_{ix}$ value which maximizes the counts, while  we have scanned the signal-mode fiber tip, also with a $200\mu$m diameter,  around the diametrically-opposed point, within the square indicated in Fig.~\ref{Fig:z0}(a).   The plot in  Fig.~\ref{Fig:z0}(b) corresponds to the number of signal-idler measured coincidence counts as a function of the position (transverse momentum) of the signal-mode fiber tip.      This plot shows, as expected from Eq.~(\ref{E:state}), an annular structure which ideally would be identical to the pump angular spectrum, plotted in Fig.~\ref{Fig:pump}(a).    The much larger width of the signal-mode angular spectrum annulus, compared to that of the pump (leading to $w_0 \approx 260\mu$m vs $1850\mu$m for the pump),
is a consequence of the considerable idler-mode fiber tip width of $200\mu$m;  the observed annulus is given by the incoherent addition of the contributions for each $\textbf{k}^\bot_i$ within the angular acceptance of the idler detector.   In addition, the width of the observed annulus is enlarged due to the convolution of the actual annulus with the angular aperture of the (signal-mode) scanning fiber tip.  Note that the  $k_t$ value that could be inferred from the radius of the the most intense region of the annulus  in Fig.~\ref{Fig:z0}(b) corresponds well to that inferred from the pump angular spectrum ($k_t = 0.046 \pm 0.001 \mu$m$^{-1}$).

Besides the conditional angular spectrum of the signal photon, shown in Fig.~\ref{Fig:z0}(b), we are interested in measuring the transverse single-photon intensity in position space.   For this purpose, we have placed a second $f$-$f$ optical system following FP1.    Specifically, we have placed a lens (focal length $f_2=30$cm and  2-inch diameter; L2) at a distance of $30$cm from FP1, so as to define a second Fourier plane (FP2) a distance of $30$cm from the lens (see Fig.~\ref{Fig:setup});   the larger lens diameter is intended to minimize signal-mode $k$-vector clipping by the lens aperture.   For this measurement,  we retain the fixed conditioning idler detector on FP1 and we place the signal-mode fiber tip with a $50\mu$m diameter, rather than $200\mu$m as for FP1,  and scan its transverse position while monitoring signal-idler coincidence counts.     The resulting data, i.e coincidence counts between idler photons collected on FP1 and signal photons collected on FP2 as a function of the position of the signal-mode fiber tip; see Fig.~\ref{Fig:z0}(c), constitutes a measurement of the heralded  singal-mode intensity as a function of transverse \emph{position} rather than \emph{momentum}.

Note that the non-diffracting propagation distance $z_{max}$ is proportional to $w_0$ which is in turn inversely proportional to the annulus width $\delta k$.  Thus, the increased width of the signal-photon conditional angular spectrum of Fig.~\ref{Fig:z0}(b) (relative to that of the pump angular spectrum),  controlled by the angular acceptance width of the idler detector, has the effect of reducing $z_{max}$ with respect to its ideal value obtained for an idealized point-like idler detector.  While a smaller-diameter fiber leading to the idler detector would enhance $z_{max}$, this unfortunately leads to a prohibitive reduction of coincidence counts. 

\begin{figure}[ht]
\centering
\includegraphics[width=8cm]{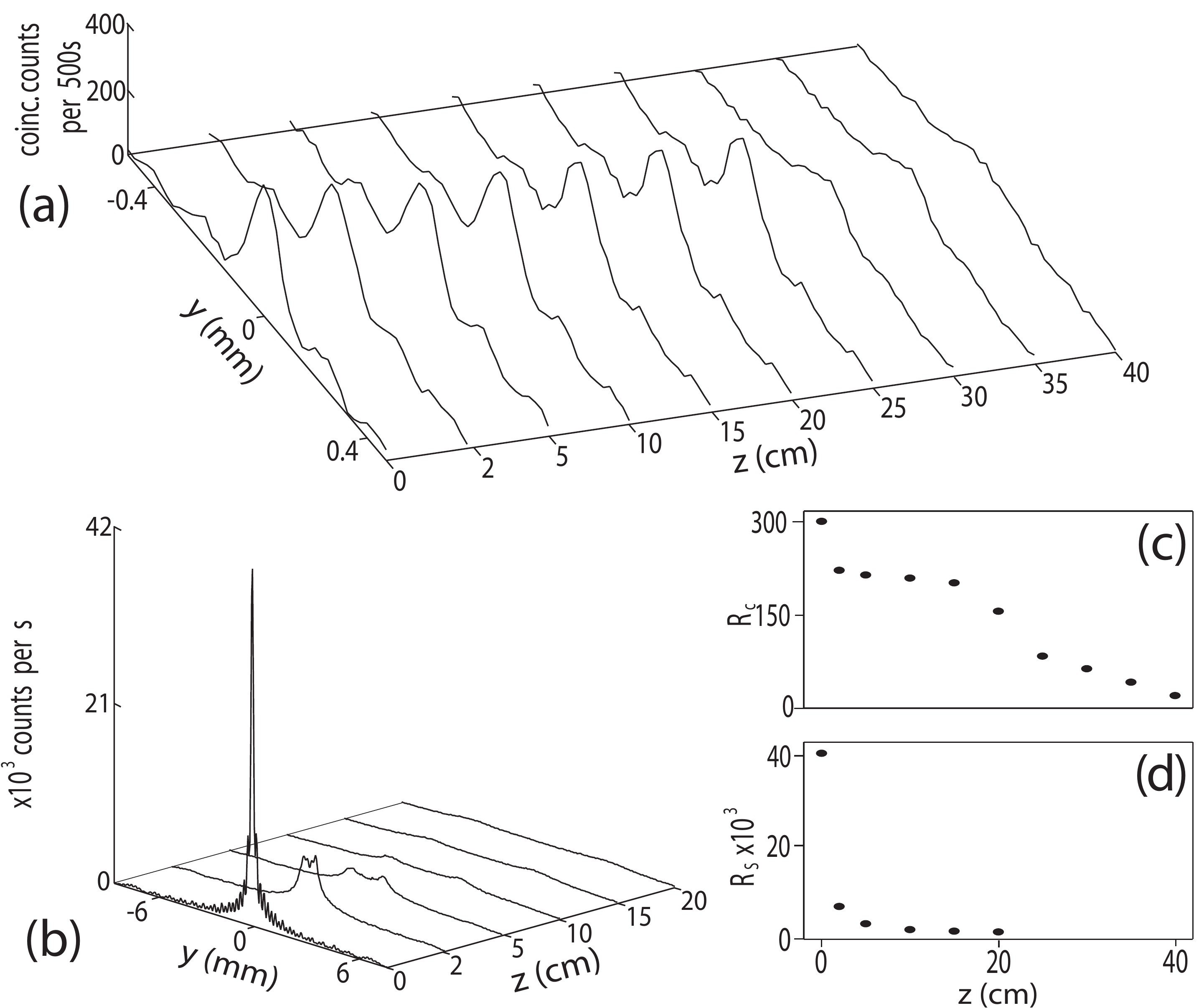}
\caption{a) Measured heralded single-photon intensity, as a function of $y$, under propagation along $z$. b) Similar to a) except showing single-channel signal counts. c) Maximum counts $R_c$ for each curve in a).  d) Maximum counts $R_s$ for each curve in b). }\label{Fig:propagation}
\end{figure}

In order to study the propagation properties of our BG signal photon we have displaced the plane on which the signal-mode fiber tip is scanned, from its initial position corresponding to the plane FP2 (regarded as $z=0$), and have collected signal-idler coincidence data for each propagation plane.   In Fig.~\ref{Fig:propagation}(a) we show the resulting signal-idler coincidences  as a function of the $y$ coordinate, for  ten propagation planes, along $z$, covering a propagation distance of $40$cm.    It can be appreciated that the width of the signal-mode transverse intensity remains essentially unchanged over a propagation distance surpassing $25$cm, with the maximum number of coincidence counts exhibiting a gradual decline.   This behavior can be directly contrasted with the corresponding behavior for single-channel counts, i.e. signal-mode detection events unconditioned by the detection of an idler photon.  Figure~\ref{Fig:propagation}(b) shows single-channel counts, as a function of $z$ and $y$, in direct correspondence with Fig.~\ref{Fig:propagation}(a).    It may be appreciated that unlike for the coincidence counts, the single-channel counts profile broadens and declines in height dramatically with propagation distance.     In a related paper, Ref.~\cite{dilorenzo09}  reports near-field, specifically intra-crystal, signal-idler correlations  which can have a variety of shapes, including a structure that resembles that of a Bessel beam for a specific value of the phase mismatch and for an unstructured pump.

The above behavior for the coincidence-counts transverse profile, under propagation along $z$, constitutes clear evidence of non-diffracting behavior  at the
single-photon level, for the heralded signal mode.  Note that this  behavior is in line with the value for $z_{max}=40.6$cm, obtained for our BG single photons through the expression $z_{max}=(f_2/f_1)^2w_0 k_s /k_t.$      Figure~\ref{Fig:propagation}(c) shows the maximum number of coincidence counts, as a function of $z$, showing a gradual decline with propagation.    Likewise, Fig.~\ref{Fig:propagation}(d) shows the corresponding behavior for the single-channel signal counts, showing a much more drastic decline with propagation.  It is interesting to note that these behaviors imply that the ratio of coincidences to single-channel counts increases by one order of magnitude over the propagation distance considered.

\section{Conclusions}

We have presented an experiment in which an SPDC nonlinear crystal is pumped by a Bessel-Gauss beam.        Our measurement of the signal-mode conditional angular spectrum shows that it corresponds, as limited by the angular acceptance of the idler detector, to the angular spectrum of the pump.    We have presented experimental data which shows that the width of the signal-mode, single-photon transverse intensity remains essentially unchanged over a distance exceeding $25$cm while, in contrast, the unconditioned signal-mode counts exhibit a drastic broadening within a few cm of propagation.  In this paper we have thus shown that the non-diffracting behavior of the pump is directly mapped to the heralded signal-mode single photons.

\section*{Acknowledgments}
This work was supported in part by CONACYT, Mexico, and by DGAPA, UNAM.

\end{document}